\documentstyle[twocolumn,prl,aps]{revtex}
\input epsf

\draft

\begin{document}

\title{Observation of total omnidirectional reflection from\\a one-dimensional
dielectric lattice~\thanks{To be published in Applied Physics A.}}

\author{D.~N. Chigrin}
\address{Universit{\" a}t Gesamthochschule Essen, Fachbereich Physik,
45117 Essen, Germany}

\author{A.~V. Lavrinenko}
\address{Belarusian State University, Department of  Physics,
Fr. Skarina Ave. 4, 220080 Minsk, Belarus}

\author{D.~A. Yarotsky, and S.~V. Gaponenko}
\address{Institute of Molecular and Atomic Physics, National Academy of
Sciences, 220072 Minsk, Belarus}

\author{\parbox[t]{5.5in}{\small
We show that under certain conditions one-dimensional dielectric lattice possesses total omnidirectional reflection of incident light. The predictions are verified experimentally using Na$_3$AlF$_6$/ZnSe multilayer structure developed by means of standard optical technology. The structure was found to exhibit reflection coefficient more then 99\% in the range of incident angles 0--86$^\circ$ at the wavelength of 632.8 nm for s-polarization. The results are believed to stimulate new experiments on photonic crystals and controlled spontaneous emission.
}}
\maketitle
\normalsize

The concept of photonic crystals is one of the challenging issues in modern condensed matter and optical science. Photonic crystals are periodic dielectric structures which offer a possibility to build up photon density of states in a similar way to electron density of states in conventional solids. Since the first pioneering works on this subject~\cite{Bykov1972,Yablonovitch1987,John1987} a great progress has been achieved in calculation of photonic band structures of one- (1D), two- (2D), and three-dimensional (3D) photonic crystals and a lot of novel applications in photonic devices have been proposed (see \cite{Joannopoulos1995,Pendry1994,Haus1994} and refs. therein). In particular, complete 3D photonic ban gap can be developed in a properly designed 3D photonic crystal. Spontaneous emission of atoms and molecules will be inhibited in such a crystal. A semi-infinite photonic crystal will exhibit total omnidirectional reflection of incident light. These issues are of great scientific and practical importance. For the optical range, within which the main applications are expected, most of experimental efforts were concentrated at 2D and 3D photonic crystals~\cite{Bogomolov1996,Bogomolov1997,Petrov1998,Gaponenko1998,Stuke1997,Vos1998,Lopez1997,Sherer1996}.  However, to get a complete 3D photonic band gap one has to build a perfect 3D dielectric lattice with the refraction index contrast equals to 2 or even higher. This still remains a serious technological problem.

In this Communication, we report that total omnidirectional reflection does not require a 2D or 3D photonic crystal~\cite{ChigrinLavrinenkoOSA}. We demonstrate theoretically and experimentally that under certain conditions a 1D photonic crystal can exhibit total reflection for all incident angles. One-dimensional photonic crystal, which  is nothing else but a well-known dielectric mirror consisting of alternating layers with a low, $n_1$, and high, $n_2$, indices of refraction, are much easier to fabricate than 3D one. Therefore the existence of a total omnidirectional reflection in a case of properly designed finite 1D dielectric lattice offers an alternative possibility to control the propagation of light.

When the plane electromagnetic wave propagates in a 1D periodic structure in the oblique with respect to the layers interfaces directions, only the normal component of the wave vector is involved in the band gap formation. The relative position of the band gap is shifted towards the higher frequencies with the internal angle and the overall forbidden gap is always closed up. Due to the loss of the degeneracy between polarizations the forbidden gaps do not coincide for the two fundamental polarizations. An obvious and common property of a 1D periodic structure follows: there is no an absolute nor complete 3D photonic band gap. In Fig.~1 the photonic band structure for a typical 1D photonic crystal with $n_1=1.2$, $\delta n=n_2/n_1=1.6$ and a filling fraction $\eta=d_2/d_1=1.0$ is shown in terms of normalized frequency $\omega \Lambda /2\pi c$ and the internal angle in the low index layer. Here $d_1$, $d_2$,  $\Lambda=d_1+d_2$, $\omega$ and $c$ are the thicknesses of the layers, the period of the structure, the frequency and the speed of light in vacuum, respectively.
The band structure has been calculated using the analytical form of the dispersion equation (see e.g. \cite{Yeh}). The overall forbidden gaps are closed up both for s- and p-polarizations (Fig. 1). The internal angles, for which the forbidden gaps are closed up, are depicted by the vertical dotted lines. However, when electromagnetic wave illuminates the boundary of the semi-infinite 1D photonic crystal, the possible values of the internal angles are restricted by the Snell's law. The higher are refraction indices of the layers with respect to the medium outside the crystal, the narrower is a cone of internal angles. The solid line in Fig.~1 corresponds to the maximum internal angle in case of ambient medium with refraction index $n=1$. For the presented structure (Fig.~1) the overall forbidden gap is opened for all incident angles in the case of s-polarized wave and within angular aperture about $\pm 60^\circ$ in the case of p-polarized wave. Due to the Brewster effect on the interface of low and high index layers, the forbidden gap is always narrower and closed up at smaller internal angle for p-polarized wave than for s-polarized one. If the index of refraction of the ambient medium is smaller than $n_1\sin \alpha_B$, which corresponds to the Brewster angle $\alpha_B$ on the interface of low and high index layers, and the index contrast in the layers is sufficiently large, the overall forbidden gap can be opened for all incident angles both for s- and p-polarized radiation. No propagating mode are allowed in the photonic crystal for any propagating mode in the ambient medium within such forbidden gap and the total omnidirectional reflection is developed. The detailed analysis of required conditions to obtain an absolute omnidirectional total reflection with 1D photonic crystals will be published in forthcoming paper~\cite{ChigrinLavrinenkoPRL}.

To verify the predictions we fabricated a lattice consisting of 19 layers of Na$_3$AlF$_6$ ($n_1=1.34$ in a wide spectral range within the visible) and ZnSe ($n_2=$2.5--2.8 in the visible range). The thickness of each layer was equal to $d_1=d_2=90$~nm, the period of the lattice being $d_1+d_2=180$~nm. The calculated photonic band structure of the lattice is presented in Fig.~2 in terms of wavelength and incident angle. The absolute omnidirectional photonic band gap exists in the spectral range of 604.3--638.4~nm (the grey area in Fig.~2). The gap to midgap ratio is about 5.5\%. The calculated transmission spectra for s- and p-polarizations at different incident angles are depicted in Fig.~3. There are clear overlapping stopbands for both fundamental polarization at any given incident angle, that is, the total reflection occurs within wide angular aperture. For calculations of the spectra we used the characteristic matrix method~\cite{Born}.

Transmission spectra for s- and p- polarizations at different incident angles in the range of 0--60$^\circ$ were measured using a `Cary 500' spectrophotometer (Fig.~4). A good agreement with theoretically predicted spectra is obtained. From Fig.~4 one can see that for spectral range 600-700~nm the transmission coefficient is very low for both polarizations even at $60^\circ$. The absolute values of transmission for s-polarization in spectral range of 630-700~nm was less than 0.001 within $\pm 60^\circ$ aperture that corresponds to the reflection coefficient 99.9\%. To examine transmission of this structure more precisely in a wider angular range, a simple set-up consisting of a He-Ne-laser and a CCD-detector was used. This set-up allows directly determine transmission coefficient of samples at angles up to $70^\circ$. For larger angles we have to measure reflection coefficient of samples. Dependence of transmission coefficients of investigated structure for s- and p-polarized incident radiation of He-Ne-laser at 632.8~nm upon angle of incidence is presented in Fig.~5. For p-polarization circles mark directly measured transmission coefficient, and squares mark data obtained from reflection measurements. Mismatch between them can be attributed to additional reflection from substrate-air and air-ZnSe interfaces. The solid (dashed) curve in Fig.~5 depicts theoretically calculated transmission coefficients for s- (p-) polarized light, the reasonable agreement with experiment is obtained. As can be seen from Fig.~5 transmission coefficient of p-polarized radiation remains below $2\cdot 10^{-3}$ in a wide angular range. Due to the Brewster effect on air-ZnSe interface at large angles it rises up to 0.33 at $80^\circ$ and then decreases again. On the contrary, transmission of s-polarized radiation monotonously decreases with growing angle of incidence (Fig.~5). Transmission coefficients less than $10^{-5}$ are beyond the possibilities of experimental set-up used. For this reason transmitted signal at more than $60^\circ$ cannot be detected. Because of this, data points for s-polarization at these angles are not presented in Fig.~5.

To summarize, at 632.8~nm the examined structure exhibits reflection coefficient for s-polarization more than 99.5\% in the angle range of $\pm 85^\circ$. Wider angular aperture is beyond the possibility of our set-up. However, there is no doubt that reflection remains very high outside the examined angle range as well. With respect to p-polarization the reflectivity can be enhanced in structures with larger number of layers and higher refraction indices.

In conclusion, we have shown theoretically and experimentally that one-dimensional dielectric structure can possess total omnidirectional reflection of incident light. Such a structure can be developed by means of standard fabrication techniques which are routinely used in optical industry. These findings are believed to stimulate new experiments on controllable spontaneous emission of atoms, molecules and solid state microstructures in the optical range~\cite{Joannopoulos1998}.

We would like to acknowledge helpful discussions with A.M. Kapitonov.

\begin{figure}
\noindent \epsfbox{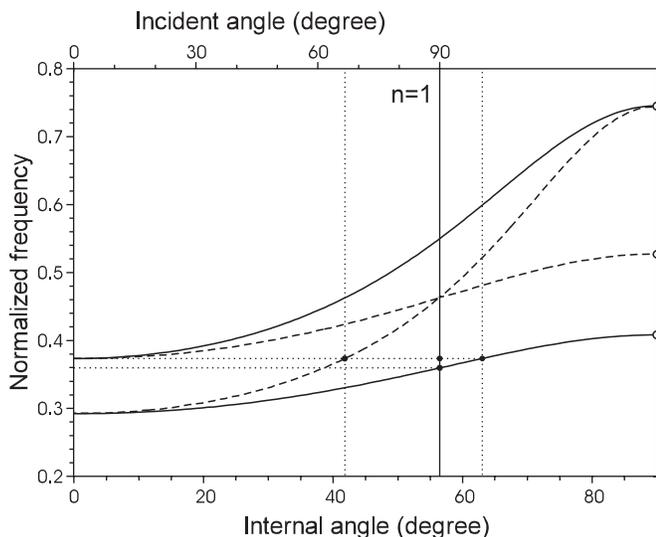}
\caption{Photonic band structure of a typical 1D photonic crystal in terms of normalized frequency $\omega \Lambda /2\pi c$ and the internal angle in the low index layer. The solid (dashed) curves are for s- (p-) polarization bands. Here, $n_1=1.2$, $\delta n=1.6$ and $\eta=1.0$.}
\end{figure}

\begin{figure}
\noindent \epsfbox{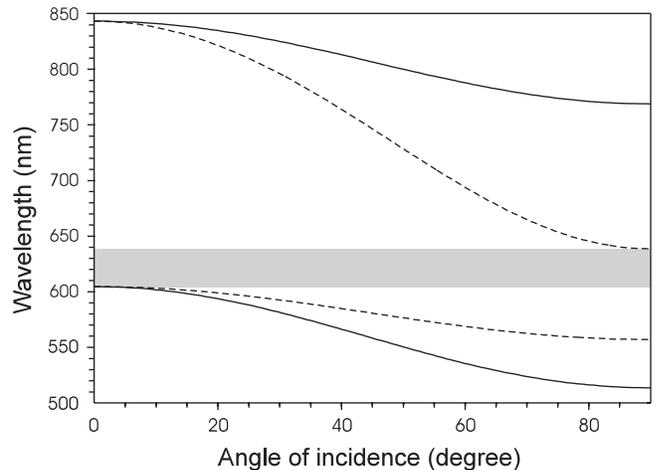}
\caption{Photonic band gap structure of semi-infinite periodic Na$_3$AlF$_6$/ZnSe lattice in term of wavelength and incident angle. The solid (dashed) curves are for s- (p-) polarization bands. The grey area is the absolute omnidirectional band gap.}
\end{figure}

\begin{figure}
\noindent \epsfbox{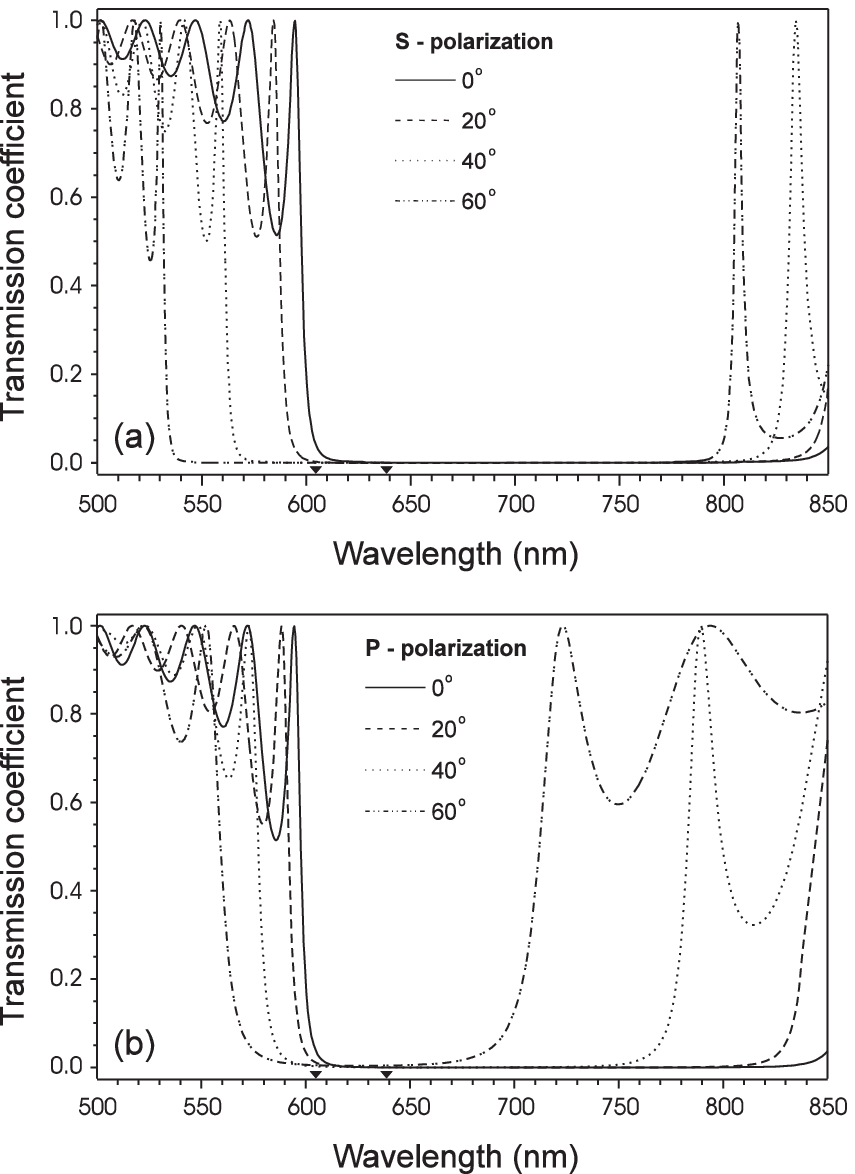}
\caption{Calculated transmission spectra of Na$_3$AlF$_6$/ZnSe 19-layer structure for s-polarized (a) and p-polarized (b) light at different angles of incidence ($0^\circ$ -- solid line, $20^\circ$ -- dashed line, $40^\circ$ -- dotted line, $60^\circ$ -- dash-dot-dotted line). The lower triangles indicate the edges of the absolute omnidirectional band gap.}
\end{figure}

\begin{figure}
\noindent \epsfbox{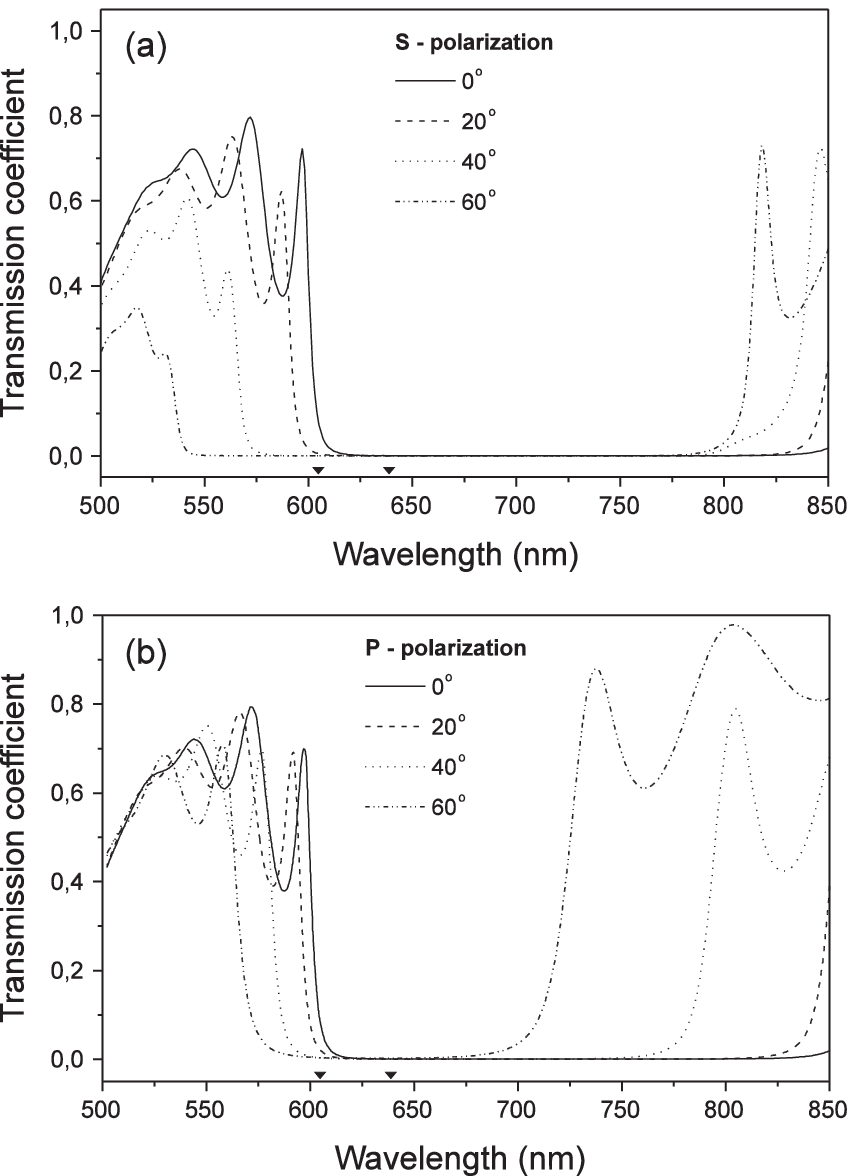}
\caption{Transmission spectra of Na$_3$AlF$_6$/ZnSe 19-layer structure measured for s-polarized (a) and p-polarized (b) light at different angles of incidence  ($0^\circ$ -- solid line, $20^\circ$ -- dashed line, $40^\circ$ -- dotted line, $60^\circ$ -- dash-dot-dotted line). The lower triangles indicate the edges of the absolute omnidirectional band gap.}
\end{figure}

\begin{figure}
\noindent \epsfbox{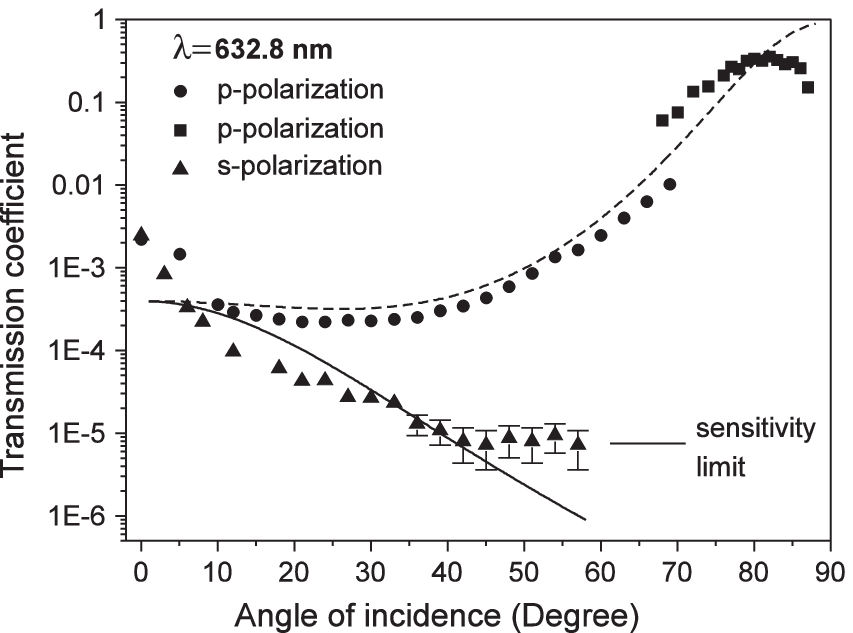}
\caption{Dependence of transmission coefficient of Na$_3$AlF$_6$/ZnSe structure upon angles of incidence at wavelength of 632.8~nm (HeNe-laser) for two polarizations of incident light. For p-polarization circles depict directly measured transmission coefficients, squares depict transmission coefficients calculated from reflection measurements data. For s-polarization (upper triangles) signal at angles more than $60^\circ$ is out of sensibility of used experimental set-up. The solid (dashed) curve depicts theoretically calculated transmission coefficients for s- (p-) polarized light.}
\end{figure}


\begin{thebibliography}{References}

\bibitem{Bykov1972} 
V.P.Bykov: JETP {\bf 62}, 505  (1972)

\bibitem{Yablonovitch1987}
E. Yablonovitch: Phys. Rev. Lett. {\bf 58}, 2059 (1987)

\bibitem{John1987}
S. John: Phys. Rev. Lett. {\bf 58}, 2486 (1987)

\bibitem{Joannopoulos1995}
J.D. Joannopoulos, R.D. Meade, and J.N. Winn:
{\em Photonic Crystals: molding the flow of light} (Princeton University Press, Princeton 1995)

\bibitem{Pendry1994}
J. Pendry: J. Mod. Opt. {\bf 41}, 209 (1994)

\bibitem{Haus1994}
J. Haus: J. Mod. Opt. {\bf 41}, 198 (1994).

\bibitem{Bogomolov1996}
V.N. Bogomolov, S.V. Gaponenko, A.M. Kapitonov, A.V. Prokofiev, A.N. Ponyavina, N.I. Silvanovich, and S.M. Samoilovich: Appl.
Phys. A {\bf 63}, 613 (1996)

\bibitem{Bogomolov1997}
V.N. Bogomolov, S.V. Gaponenko,  I.N. Germanenko, A.M.
Kapitonov, E.P. Petrov, N.V. Gaponenko, A.V. Prokofiev, A.N. Ponyavina,
N.I. Silvanovich, and S.M.Samoilovich: Phys. Rev. E {\bf 55}, 7619 (1997)

\bibitem{Petrov1998}
E.P. Petrov, V.N. Bogomolov, I.I. Kalosha, and S.V.
Gaponenko: Phys. Rev. Lett. {\bf 81}, 77 (1998).

\bibitem{Gaponenko1998}
S.V. Gaponenko, A.M. Kapitonov, V.N. Bogomolov,  A.V.
Prokofiev, A. Eychmueller, and A.Rogach: JETP Letters {\bf 68}, 142 (1998).

\bibitem{Stuke1997}
M.C. Wanke, O.Lehmann, K. M{\" u}ller, Q. Wen, and M. Stuke: Science {\bf 275}, 1284 (1997)

\bibitem{Vos1998}
J.E.G.J. Wijnhoven and W.L. Vos: Science {\bf 281}, 802 (1998)

\bibitem{Lopez1997}
R. Mayoral, J. Requena, J.S. Moya, C. Lopez, A. Cintas, H.
Migues, F. Meseguer, L. Vazquez, M. Holgado, and A. Blanco: Adv.
Materials {\bf 9}, 257 (1997)

\bibitem{Sherer1996}
R.C. Tyan, P.C. Sun, A.Scherer, and Y. Fainman: Opt. Lett. {\bf 21}, 761 (1996)

\bibitem{ChigrinLavrinenkoOSA}
D.N. Chigrin and A.V. Lavrinenko, in {\em Technical Digest of the 1998 OSA Annual Meeting and Exhibit}, (Baltimor, Maryland, USA, 1998), p. 118.

\bibitem{Yeh}
P. Yeh, {\em Optical Waves in Layered Media} (John Wiley and Sons, New York, 1988).

\bibitem{ChigrinLavrinenkoPRL}
D.N. Chigrin and A.V. Lavrinenko (in press)

\bibitem{Born}
M. Born and E. Wolf, {\em Principles of Optics} (Pergamon, New York, 1980).

\bibitem{Joannopoulos1998} When the manuscript of the present paper has been
completed, a theoretical paper by J. Winn, Y.Fink, S. Fan, and J.D.
Joannopulos was issued (Opt. Lett. {\bf 23}, 1573, October 15, 1998) which
contains basically the same idea but without experimental verification.

\end{thebibliography}
\end{document}